# Free and Trapped Injected Carriers in $C_{60}$ Crystals

V.Y. Butko* and A. P. Ramirez
Los Alamos National Laboratory, Los Alamos, New Mexico, 87545

C. Kloc
Bell Laboratories, Lucent Technologies, 600 Mountain Avenue, Murray Hill, New Jersey, 07974

Abstract

We report on the conductance from two-contact carrier injection in $C_{60}$ single crystals. In the nonlinear regime, the current and voltage obey a power law, $I \propto V^m$, where m can be as high as 10 at room temperature. This nonlinear behavior – the resistance decreases by 6 orders of magnitude without saturation – is among the highest reported for organic systems, and can be explained by injection of free carriers into the trap-filling region. We find that $H_2$ annealing suppresses shallow traps and enhances nonlinearity. Two limiting types of temperature dependence of the nonlinear resistance are observed – decreasing and increasing resistance at the orientational ordering temperature. A simple model incorporating deep traps is presented to understand this behavior and the impact of this model on possible field-effect transistor action is discussed.

Devices such as the Field Effect Transistor (FET), based on the controlled injection of free carriers into semiconductor crystals, are the fundamental devices of modern electronics. It is thought that fabricating such devices from organic semiconductors will lead to new applications where low cost and large area are desired [1]. Among the most interesting of simple organic systems is fullerite, a solid comprised of $C_{60}$ molecules arranged in a face-centered-cubic lattice. The interest stems not only from the simplicity of the $C_{60}$ – implying high mobility devices are achievable – but also from the superconductivity that results when doped with alkali metal [2]. This latter feature implies that at large densities of injected charge, a new class of superconducting devices with large on/off ratios might be achievable. The ability of a substrate to easily conduct injected carriers is critical for any of these applications. For materials possessing a high trap density, the current from injected carriers can be completely suppressed. Therefore an investigation of the carrier release and trapping processes in $C_{60}$ crystals is essential for understanding the fundamental transport mechanisms and for reproducibly fabricating FET devices from $C_{60}$. For these purposes, we studied the non-ohmic transport using the two-contact method in a standard current-voltage measurement on a number of different $C_{60}$ samples. This method is used as spectroscopic probe of the trap density [3, 4] and assumes that the current due to intrinsic carriers in the semiconductor is negligible compared to a space charge limited current (SCLC) of the injected carriers [3].

All crystals studied in this work have been grown at Bell Labs using the same physical vapor transport method as reported in Ref.[5]. All contact preparation, $H_2$ annealing, and measurements have been done at Los Alamos National Laboratory. Typical crystal dimensions are 1-2 mm length, 0.2 – 1 mm width and 0.05 - 0.5 mm thickness. For ohmic contacts, gold electrodes (~ 900A thick) were thermally evaporated through a shadow mask on the smooth, untreated single crystal surface. The separation between contacts is either 12.7 or 25 µm. The width of the contact pads is ~0.1- 0.2 mm. Some samples were annealed before measurement in an $H_2$ atmosphere for 24 hours at 170 C. The current (I) and voltage (V) were measured with a Keithley 6517A electrometer. Currents down to $10^{-14}$ A, and positive voltages up to 400 V applied between one contact and another connected to ground, were routinely obtained. The electrical measurements are made in darkness in a Quantum Design cryostat at fixed temperature. For these measurements, a 2.5 V step and a 10 second delay between each measurement was typical.

It is important that the contacts between evaporated gold and $C_{60}$ crystals is ohmic [3, 6]. To verify the presence of ohmic contacts, I-V characteristics of two contact pairs with contact separation 12.7 and 25 μm on the same crystal (#19) have been measured, and the results are shown in the inset of Fig. 1. The observed difference between two measured curves is consistent with ohmic contacts and a non-linear relationship between the effective resistance and channel length [3]. For the other measurements reported here, sample #18 had a contact spacing of 12.7 μm, while all other samples had a contact spacing of 25μm.

The measured room-temperature I-V characteristics for 5 different samples are shown in Fig.1. In general, three separate voltage regions can be distinguished. All samples at low voltage demonstrate nearly ohmic I-V dependence. For samples #15 and #18 before annealing, this ohmic region extended up to the highest measured voltages. In this region, increasing applied voltage, and the correspondent increase of injected carrier density does not lead to pure free-carrier behavior due to highly effective trapping processes. Clearly the presence of such trap-dominated processes will be deleterious for FET applications. In a different set of samples, #16, #17, and #19 measured before $H_2$ annealing, a second characteristic high-voltage behavior has been observed. This behavior is also observed in all samples measured after the annealing process. We find that, in this region, the current and voltage obey a power law, $I \propto V^m$, where m can be as high as 10 at room temperature. The theory of semiconductors [6] predicts a power law behavior for cases where deep traps are distributed in either a Gaussian manner or exponentially within the forbidden energy gap. (A power law with m ~ 10 is indistinguishable here from exponential behavior, but the following analysis doesn't depend critically on such a distinction.) This leads us to the conclusion that these crystals have a low density of shallow traps, and that the trap density above a certain threshold decreases sharply with energy. In the case of electron traps with a Gaussian distribution, deep and shallow means $E_v < E_{tm} < E_F$ and $E_c > E_{tm} > E_F$ respectively, where $E_v$ is the valence band, $E_{tm}$ is the centroid of the distribution, $E_F$ is the Fermi energy, and $E_c$ is the conduction band [6]. A similar definition applies for holes.

The third characteristic voltage region obeys a Child's power law for space charge limited conduction (SCLC) [3],

$$J = \frac{9}{8}\mu\varepsilon\varepsilon_0 \frac{V^2}{L^3} \quad , \qquad (1)$$

and has been observed in only one of the samples studied here (#16). Here J is the current density, μ is carrier mobility, ε is dielectric constant of the semiconductor ($\varepsilon \sim 4.4$ for $C_{60}$ [7]), $\varepsilon_0$ is permittivity in vacuum, V is the applied voltage, and L is the distance between contacts. In this region, using an approach described in Ref. [3], the carrier mobility can be estimated from equation (1) as 0.4 $cm^2/(V*s)$ at room temperature. Here we have arbitrarily assumed that current flows only through a surface layer with thickness of 20% of the distance between the contacts (according to equation (1) J is proportional $1/L^3$ and quickly decreases with length of the current path). This rough mobility estimate does not take into account any crystalline anisotropy but gives a reasonable value which is smaller by only a factor 3-5 times the highest mobility values determined from time-of-flight and FET measurements [8-10].

As shown in Fig. 1, the main effect of annealing is to increase the exponent, m, in the region of non-linear region. We use the following general theoretical consideration to describe the hydrogen annealing effect on the trap density, assuming both electron and hole transport is possible in $C_{60}$. We then assume that annealing primarily affects the carrier transport by changing the trap density, without affecting the intrinsic carrier mobility. The current density (J) is given by

$$J = e\mu n_f V \,, \qquad (2)$$

where e is the electron charge, μ is the mobility, $n_f$ the density of free carriers, V is the applied voltage [6].

In the SCLC region, the average total injected bulk charge (Q) [3] is given by an expression: $Q \approx CV$, where C is the capacitance. Therefore the total average number of electrons injected in the bulk is a function only of the dielectric constant, the geometrical parameters of the measured 2-contact structure, and the applied voltage, all of which should not significantly changed by the annealing. Assuming that all injected electrons are ether trapped or free, we can write for the average densities of free ($n_f$), trapped, ($n_t$) and total injected electrons (n) before and after annealing,

$$n_{fo} = n - n_{to} \qquad (3)$$

$$n_{fa} = n - n_{ta} \,. \qquad (4)$$

From (2), (3), and (4) we get the following equation

$$\frac{J_o}{J_a} = \frac{n_{fo}}{n_{fa}} = \frac{n - n_{to}}{n - n_{ta}}. \quad (5)$$

For $J_a >> J_o$, using equations (5) and (2), we obtain

$$\frac{d(n_{to} - n_{ta})}{dV} \propto V^{m_a - 1} \quad (6)$$

For sample #19 in the trap-filling region $m_a$ reaches 10. Therefore, we find that the primary effect of annealing is to suppress shallow traps at higher voltages.

This conclusion can be clarified from the data for sample #15. For this sample, the trap-filling region was observed only after annealing, but the current $J_a < J_o$ at all measured voltages. This behavior can be explained by the assumption that annealing not only suppresses traps but also provides a trap distribution shift to low voltages, i.e., by a shift of trap spectral weight to lower energies for a single peak in the trap density, so that the number of trapped carriers can increase in spite of decreasing of total number of traps. Therefore we see that these results demonstrate an effective room-temperature injection of free carriers into $C_{60}$ single crystals and this process is enhanced greatly after annealing in hydrogen. This behavior is a potential advantage for making $C_{60}$ FET structures with high on/off ratios.

Another essential requirement for making superconducting FETs from $C_{60}$ is an effective free carrier injection at temperatures low enough for carrier coupling. To analyze this condition we have extended the I-V measurements below room temperature. We find two limiting types of behavior associated with the orientational ordering in the vicinity of 260K. For the first type of behavior, shown in Fig. 2, the effective resistance versus voltage at all temperatures is qualitatively similar to the room temperature behavior, namely a nearly ohmic region followed by a strongly nonlinear trap-filling region, as shown in Figs. 2. For this behavior, which we call deep trap carrier transport (DTCT), the slope of the I-V curves in the nonlinear region increases monotonically with decreasing temperature, as shown in the inset of figure 2. This effect can be explained by a sharper carrier energy distribution at low temperatures and the resulting faster increase of free carrier density with voltage. This result is also qualitatively consistent with the theoretical prediction for the deep traps $m = \sqrt{1 + 2\pi\sigma_t^2 / 16k_B^2T^2}$, where $\sigma_t$ is the width of the trap density of states peak [6]. A similar dependence was reported in studies of polymer films [11]. However, the absolute values of the slope, as well as the low-voltage resistances for the $C_{60}$ crystals studied here are significantly higher. However, the most important feature of the DTCT type of behavior is the nearly monotonic increase of resistance, in the nonlinear region, with decreasing temperature. As one can see in both Fig. 2 and in the top part of Fig. 4, this temperature dependence is significantly sharper in the region of orientational ordering 240-260 K [12]. The second curve of Fig. 2 inset demonstrates that the temperature dependence of the voltage at fixed current ($3\times10^{-11}$ A) also has a well-defined transition shape in the critical region, 240-260 K.

In Fig.3 is shown the second type of limiting behavior. This behavior is characterized by a non-monotonic temperature dependence of the nonlinear resistance. Near room temperature and at the lowest temperatures the R(V) curves are nested and have a quasi-ohmic behavior followed by the power law nonlinear region. In the intermediate temperature range and especially at the orientational ordering temperatures the (V) dependence is more complicated and does not obey a power law. In particular, the several-order-of-magnitude drop of the resistance with temperature in the temperature region surrounding the orientational ordering is very unusual for semiconductors. This type of behavior is qualitatively similar to previous results of the temperature dependence of the ohmic resistance [13, 14]. The temperature dependence of the resistance at fixed voltages in the nonlinear region is presented in the bottom part of Fig. 4. This type of behavior can be explained by a combination of DTCT and another mechanism which we call shallow trap carrier transport (STCT).

We present a simple model to explain both types of temperature dependences of the non-linear resistance in the case of single-carrier injection [6]. We assume that actual trap density distribution within the forbidden energy gap can be represented by a superposition of one or more Gaussian peaks and that the Fermi energy monotonically decreases with decreasing temperature. For example, for both samples exhibiting STCT in Fig.4, the two resistance peaks corresponded to two peaks in the trap density-of-states within this model. In the top part of the Fig.3, for purposes of simplicity, only one peak has been plotted, and only for the case of electron transport. As was mentioned above, the theory of semiconductors predicts a power law behavior for deep traps distributed in a Gaussian manner within the forbidden energy gap [6]. The density of states, h(E), is given by,

$$h(E) = A \exp\left(\frac{(E - E_{tm})^2}{2\sigma_t^2}\right), \qquad (7)$$

where A is an energy-independent factor. Here and below we consider average values of the carrier and trap density in the energy region of overlap.

For electron DTCT behavior, the Fermi energy ($E_{Fn}$) decreases with temperature but in the investigated temperature range $E_{Fn} > E_{tmn}$ (see the top part of Fig.3.) Decreasing temperature in this case decreases the free carrier number primarily due to a decrease of the thermal activation term.

As in Ref. [4] we assume that at any injection rate, thermal equilibrium is established between free and trapped carriers. The density of free carriers can thus be described in terms of conventional Boltzmann statistics.

$$n_f = N_e \exp\left(-\frac{E_c - E_{Fn}}{k_B T}\right) \qquad (8)$$

where $N_e$ is the effective density of states. The factor $N_e$ in the case of SCLC is proportional to the difference between average densities of the injected and trapped carriers $(n_e - n_t)$ see equations (3) and (4).

$$N_e \propto (n_e - n_t) \qquad (9)$$

The density of the trapped carriers can be estimated by using expression (7) and assuming that all traps below the Fermi level are filled and that all traps above the Fermi energy are empty. This approach was used in Ref. [4].

$$n_t = \int_0^\infty dEh(E)H(E - E_{Fn}) \approx \int_0^{E_{Fa}} dEh(E) = Bh(E_{Fn}) \qquad (10)$$

Where $H(E\text{-}E_F)$ is the Heaviside step function, B is a factor almost independent of energy. From equations (8), (9) and (10) we obtain,

$$n_f = (n_e - Bh(E_F))\exp\left(\frac{E_c - E_{Fn}}{k_B T}\right) \qquad (11)$$

For the second type of behavior the Fermi energy, as usual, decreases with decreasing temperature and this effect can change the relative values of $E_{tmn}$ and $E_{Fn}$ from the relation $E_{Fn} > E_{tmn}$ (DTCT) to $E_{tmn} > E_{Fn}$ (STCT). For $E_{tmn} > E_{Fn}$ the absolute value of the second term in equation (11) can decrease sharply with decreasing temperature. This effect is based on increased trapped carrier release with decreasing temperature. This process, which is valid for both holes and electrons, can result in an increase of free carrier density and, therefore, a decrease of resistance.

In the above model, therefore, a decrease in Fermi energy with temperature can result in changing the relative values of $E_{tm}$ and $E_F$ and thus lead to non-monotonic temperature dependence, as observed for the second type of behavior. In this model the difference between the first and second type of behavior can be explained by a broader peak distribution in the first type or by a different initial (room temperature) values of $E_{tm}$ and $E_F$.

For the R(V) dependence of sample #15, shown in the Fig. 3, at the lowest temperatures, the above model suggests that the Fermi energy decreases and encounters another peak in the trap density of states, and this situation can be qualitatively described by the same deep trap distribution theory as for samples of the first type, and also for sample #15 at room temperature. For two samples (#19 and #15) we have been able to measure I-V characteristics down to 40 K and 90 K respectively and still observe a trap-filled region, demonstrating that effective carrier injection can be realized at low temperature.

An opposite sign of the temperature dependence in the phase transition region for the different samples implies that the orientational ordering does not just change the density of traps, but shifts the trap distribution. This shift can either increase or decrease the number of deep traps. Therefore, the sign of the temperature dependence depends on the initial relative position of the Fermi energy and the trap distribution centroid. The physical nature of the traps in $C_{60}$ crystals is still open question. One possibility is that frozen fluctuations in the orientational order parameter act as carrier traps.

In conclusion, we have found that $C_{60}$ crystals demonstrate distinct types nonlinear transport behavior. For most of the samples at high enough temperatures, a nonlinear region has been observed. Meanwhile some samples before annealing do not show nonlinear I-V behavior at room temperatures and FETs made from such samples will probably perform poorly. Two limiting types of temperature dependence of the resistance are observed – decreasing and increasing resistance in the intermediate temperature range and particularly at the orientational phase transition. This observation suggests that in a limited temperature range, lower voltages are required to inject charge at low temperatures than at high temperatures. Another finding of high practical importance is that annealing $C_{60}$ crystals in hydrogen effectively suppresses traps, shifting the trap distribution to lower energies. This procedure is thus recommended for most FET applications. Extrapolation of our low temperature results for annealed $C_{60}$ crystals makes it reasonable to believe that, for 20-30% of the samples, effective free carrier injection can be realized in the temperature range where superconducting carrier coupling is possible.

We are grateful to Chandra Varma, Gavin Lawes, and Charley Chi for help and useful conversations, and we acknowledge support from the Laboratory Directed Research and Development Program at Los Alamos National Laboratory, and the Nanoscience Engineering and Technology Program at the United States Department of Energy Office of Science.

* On leave from Ioffe Physical Technical Institute (PTI), Russian Academy of Science, Polytechnicheskaya street, 26, St. Petersburg, Russia

**Figure Captions**

1. Room temperature current voltage (I-V) characteristics. Curves $C_{60}$ #15, $C_{60}$ #16, $C_{60}$ #17, $C_{60}$ #18 and $C_{60}$ #19 correspond to the samples #15, #16, #17, #18and #19 before annealing. Curves $C_{60}$ #15-an, $C_{60}$ #16-an, $C_{60}$ #17-an, $C_{60}$ #18-an and $C_{60}$ #19-an correspond to samples #15, #16, #17, #18and #19 after the annealing. Room temperature I-V characteristics of two contact pairs with contact separation 25 µm and 12.7 µm on the same crystal (#19) in the inset.

2. The voltage dependence of resistance at different temperatures for sample #19. This figure illustrates type I temperature behavior and the phase transition at 250-270K. The black curve of the inset demonstrates the temperature dependence of the voltage at fixed current ($3\times10^{-11}$ A). The slope of the I-V curves in the nonlinear region increases monotonically with decreasing temperature corresponds to the red curve in the inset.

3. The bottom part of the figure presents voltage dependence of resistance at different set of temperatures for sample # 15. The top part is a schematic illustrating energetic trap distribution within the forbidden energy gap and relative positions of an electron Fermi energy and the centroid of the distribution for LTE and HTE samples.

4. The top part of the figure shows the temperature dependence of resistance for LTE samples. Curves #12, 351 V and #12, 276 V correspond to sample #12 measured at 351 and 276 V, respectively. Curves #17, 351 V and 19, 211V correspond to sample #17 measured at 351 and sample #19 measured at 276 V, respectively. The bottom part of the figure shows a temperature dependence of resistance for HTE samples. Curves #15, 349 V and #15, 201 V correspond to sample #15 measured at 349 and 201 V, respectively. Curve #18, 176 V corresponds to sample #18 measured at 176 V.

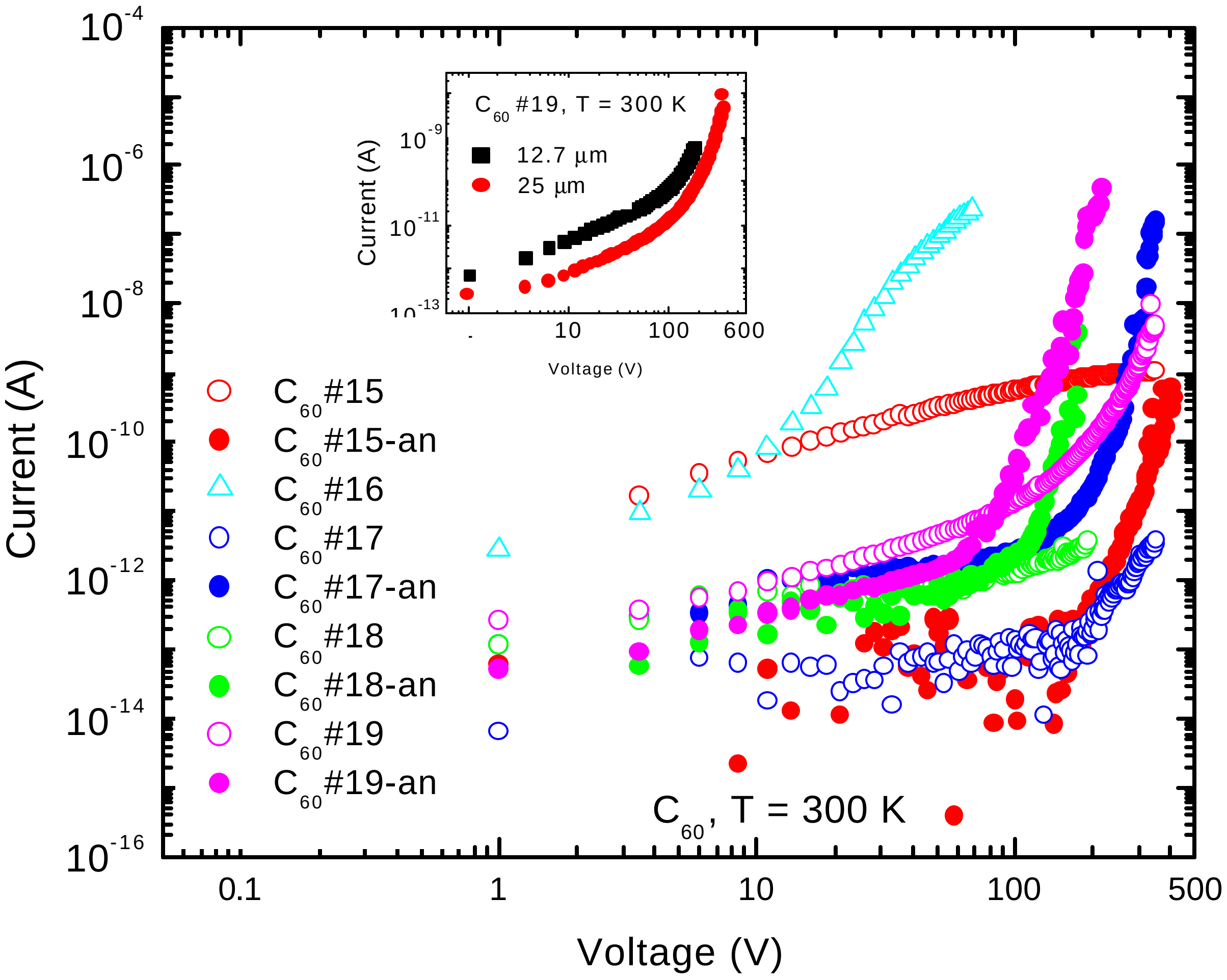

Fig. 1

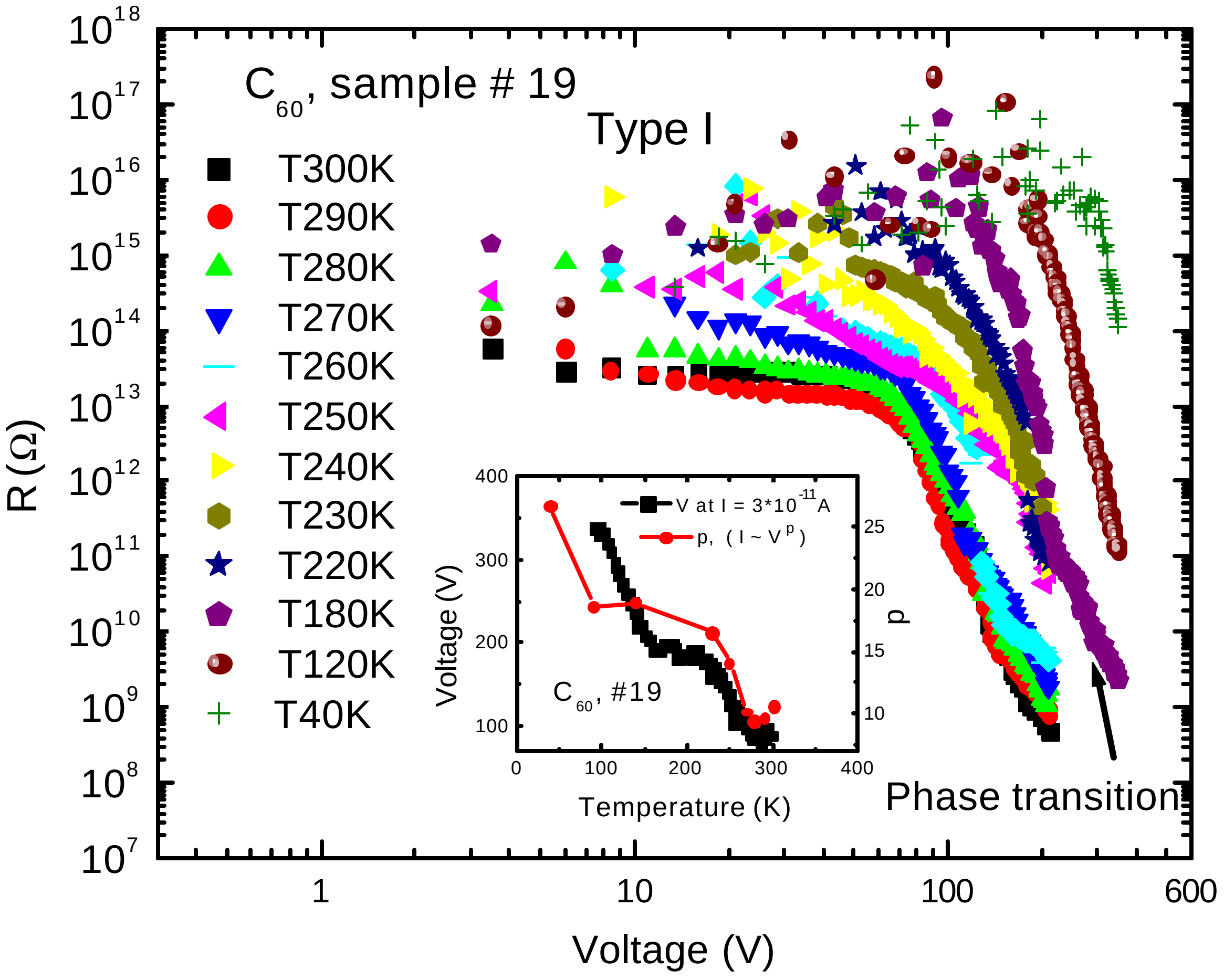

Fig. 2

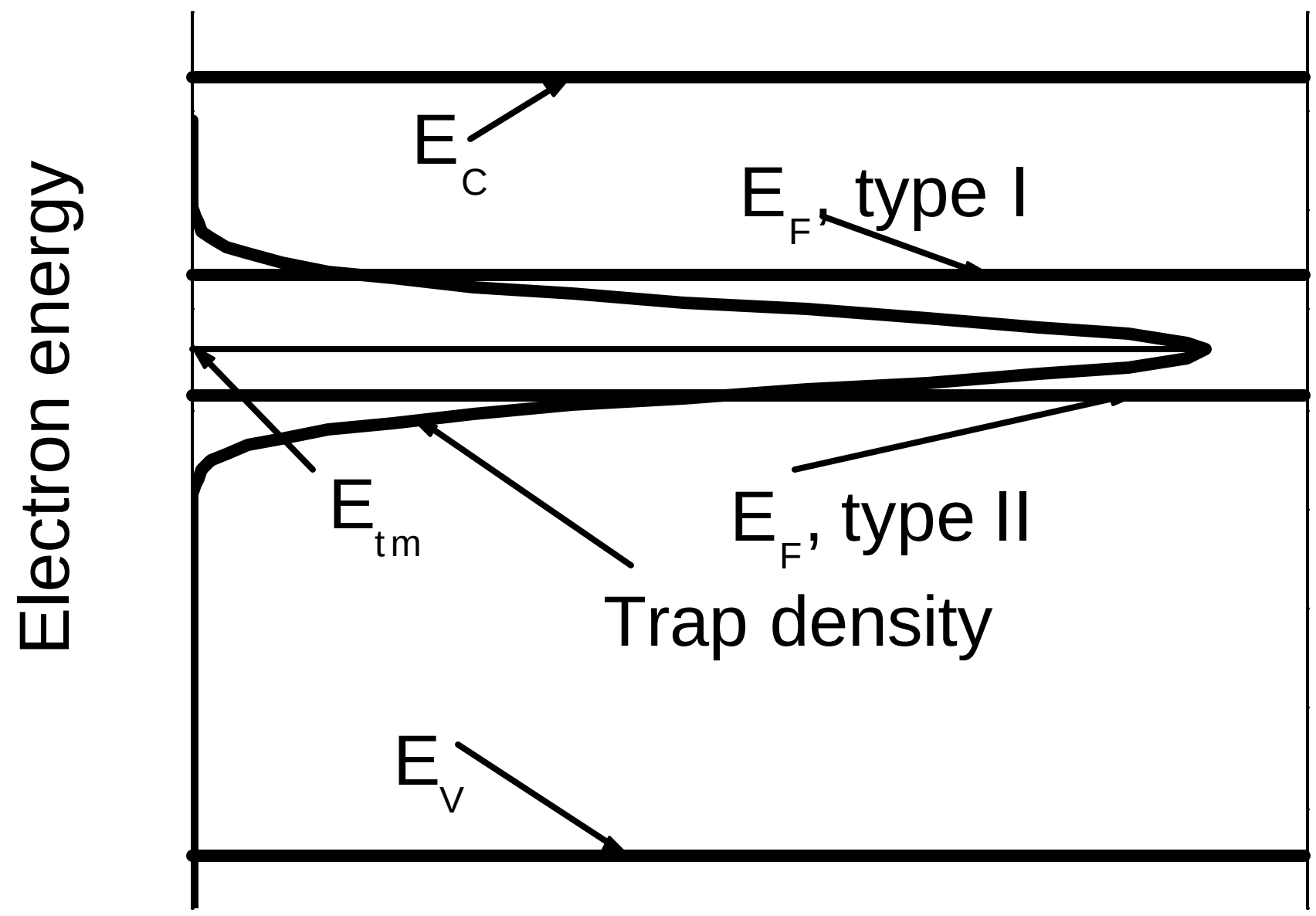

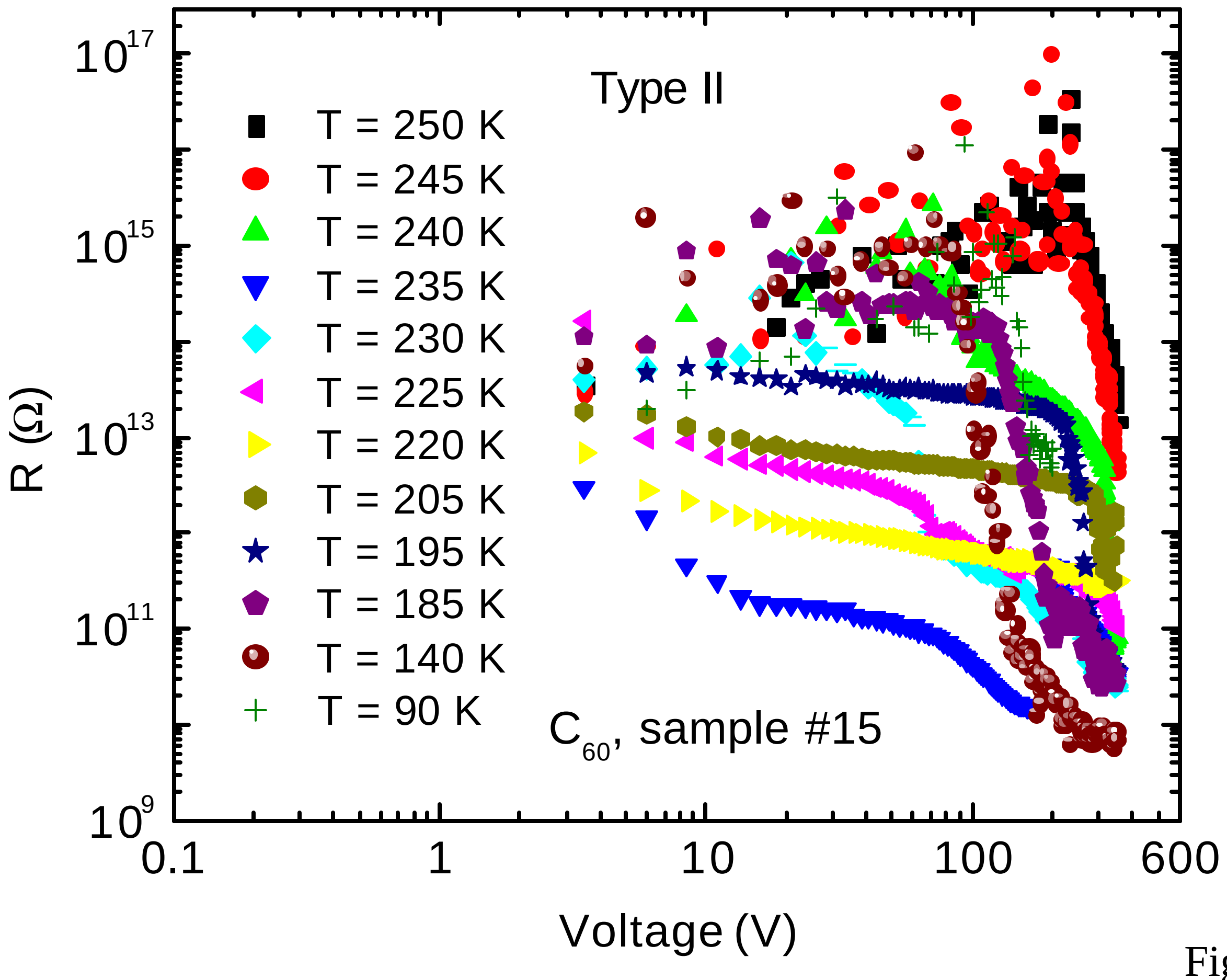

Type II
T = 250 K
T = 245 K
T = 240 K
T = 235 K
T = 230 K
T = 225 K
T = 220 K
T = 205 K
T = 195 K
T = 185 K
T = 140 K
T = 90 K
C60, sample #15
R (Ω)
10^17
10^15
10^13
10^11
10^9
0.1
1
10
100
600
Voltage (V)

Fig. 3

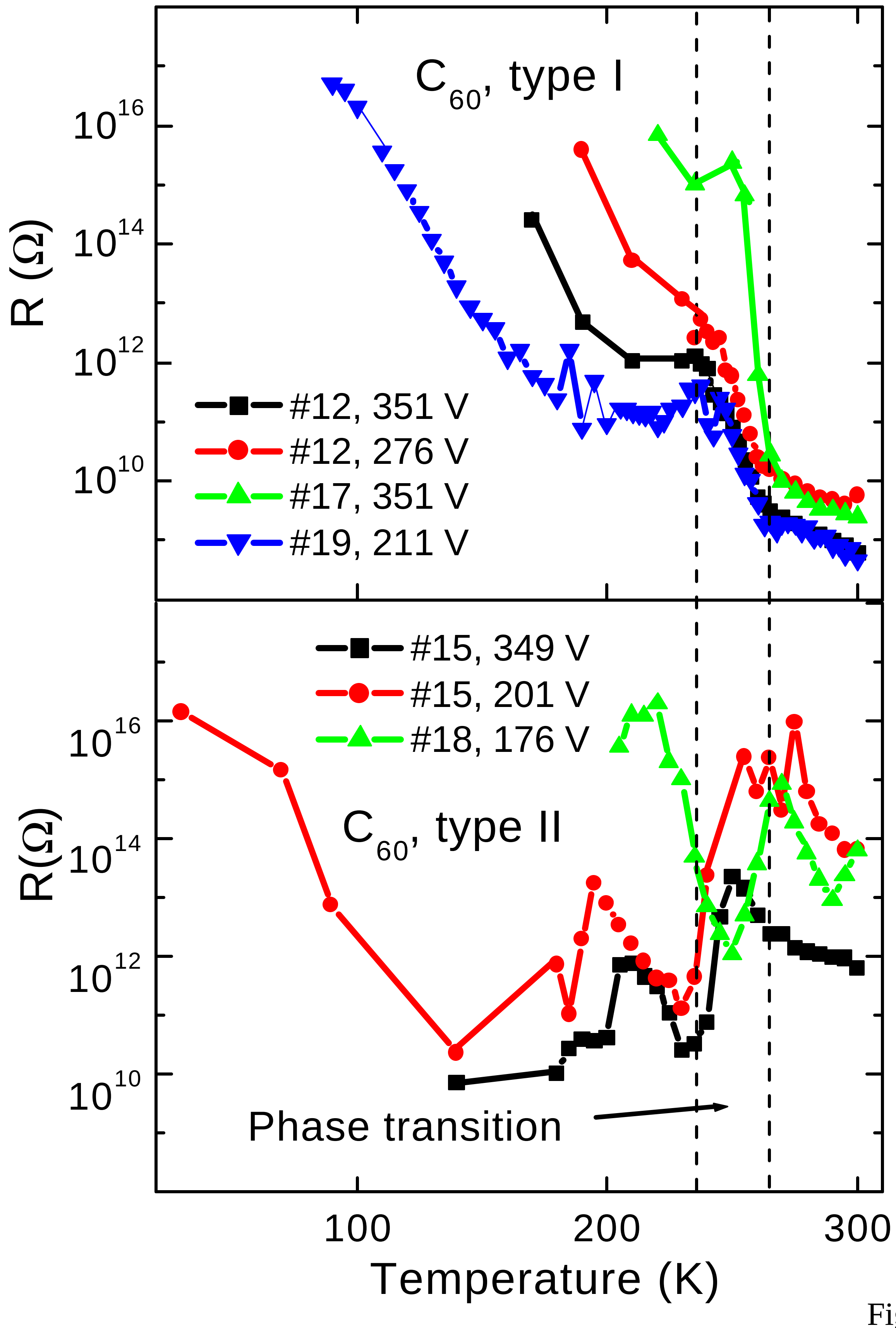

Fig. 4